\documentclass[3p,times]{elsarticle}

\usepackage{ecrc}

\volume{00}

\firstpage{1}

\journalname{Nuclear Physics A}

\runauth{}

\jid{nupha}

\usepackage{amssymb}

\begin{document}

\begin{frontmatter}



\dochead{}

\title{Analyzing the Power Spectrum of the Little Bangs}


\author[a]{\'Agnes M\'ocsy}
\author[b]{Paul Sorensen}
\address[a]{Pratt Institute,  Department of Math and Science, Brooklyn, NY 11205 USA}
\address[b]{Physics Department, Brookhaven National Laboratory, Upton NY 11973 USA}
\begin{abstract}
In this talk we discuss the analogy between data from heavy-ion collisions and the Cosmic Microwave Background. We identify $p_T$ correlations data as the heavy-ion analogy to the CMB and extract a power-spectrum from the heavy-ion data. We define the ratio of the final state power-spectrum to the initial coordinate-space eccentricity as the transfer-function. From the transfer-function we find that higher $n$ terms are suppressed and we argue that the suppression provides information on length scales like the mean-free-path. We make a rough estimate of the mean-free-path and find that it is larger than estimates based on the centrality dependence of $v_2$.
\end{abstract}


\end{frontmatter}



A commonly quoted goal of the heavy-ion programs at Brookhaven National Laboratory and CERN is to recreate conditions similar to those shortly after the Big Bang~\cite{hatsuda}. Learning about the early stages of the matter produced in heavy-ion collisions from the observation of hadrons in the final state is in some sense similar to understanding the stages of the early universe from the observation of the Cosmic Microwave Background (CMB)~\cite{cmb}. In this talk we explore the analogy between heavy-ion collisions and Big Bang cosmology~\cite{mishra}. In particular, we present the heavy-ion equivalent of the CMB and from that we determine the power-spectrum for the little bangs. We then estimate the transfer-function necessary to produce the spectrum from the initial conditions of the collisions.

Quantum fluctuations from the early universe show up as hotspots at the surface of last scattering, creating the CMB measured for example by WMAP. Measurements of the CMB reveal temperature fluctuations corresponding to over- or under-densities present at the surface of last scattering at about 400,000 years after the Big Bang~\cite{cmb}. These density fluctuations ultimately explain the structure in our universe. Maps of the temperature of the CMB are determined from measurements of the blackbody spectra at  2M points in the sky. The temperature is found to be very smooth with relative variations only showing up at approximately $10^{-5}$. The fluctuations tend to only be at short distances; small lumps not large ones. This result has been explained by an inflationary period when the larger scale fluctuations were pushed outside of the horizon, causally separating them.

The scale of the correlations can be most easily studied by extracting a power-spectrum from the CMB. The power-spectrum shows that most power is for large wave-numbers (meaning small wavelength). The lack of power at small wave-number is strong evidence of inflation. Peaks in the power-spectrum are caused by acoustic phenomena in the early universe as density perturbations in the universe propagate as sound waves. This gives rise to the structure in the anisotropies of the microwave background, notably a characteristic angular scale and the famous acoustic peaks. From this spectrum the acoustic peaks are fit with models to extract cosmological parameters. In this talk we explore an analogous measurement in heavy-ion collisions. We determine the power-spectrum for the little bangs and we estimate the transfer-function necessary to produce the spectrum from the initial conditions.

Just as quantum fluctuations stretched to cosmic sizes by inflation show up in the CMB, we expect fluctuations from the beginning of the little bangs to show up in heavy-ion data. Mishra et.~al first have explored the analogy between the expansion of heavy-ion collisions starting from a lumpy initial energy density and the expansion of the universe starting with quantum fluctuations stretched to cosmological sizes, and based on that analogy suggested measuring RMS values of the Fourier coefficients $v_n$ (known as flow coefficients) as the power-spectrum~\cite{mishra}. They didn't, however, make a connection between the RMS of $v_n$ and the already existing two-particle correlation measurements. Sorensen later pointed out~\cite{sorensen1} that many of the surprising correlation structures seen in RHIC data (the ridge and mach cones~\cite{data}) may be understood in terms of fluctuations of higher Fourier components of $v_n$ which Mishra et. al. had argued would  arise from anisotropies in the initial energy density converted into momentum-space during the expansion. Several other authors have also previously explored the connection between initial conditions and two-particle correlations~\cite{voloshin,gavin,mclerran,brazil,v3,hannah}.

While quantum fluctuations stretched to cosmic sizes by inflation explain the temperature fluctuations in the CMB, letÕs see how density fluctuations at the beginning of heavy-ion collisions might show up in the correlations data. We have some expectations about the initial state: given that the collisions involve nuclei with a finite number of nucleons, we expect  that the initial density is inhomogeneous, it has a lumpiness. Every event is unique and no event is homogeneous. We do not dwell for the moment on the exact theory to describe these initial fluctuations, but only point out that the finite number of charges in the approaching nuclei implies that the initial (energy) density is lumpy (even if we consider a larger number of gluons in the nucleus, those gluons are still confined in the radial direction within the typical size of the nucleons). The lumpy initial density may survive through the evolution and show up at the end as hotspots in the surface of last scattering. We expect this because the rapid expansion boosts these lumps outward and the decay of the hotspots in coordinate-space manifests in correlations in momentum-space due to flow~\cite{voloshin}. The lumpy initial conditions show up in the 2-particle correlation structures. The plausibility of this conjecture has been verified with multiple models for heavy-ion collisions~\cite{brazil,v3,hannah,werner}. 

\begin{figure}
\begin{centering}
\includegraphics[width=15cm]{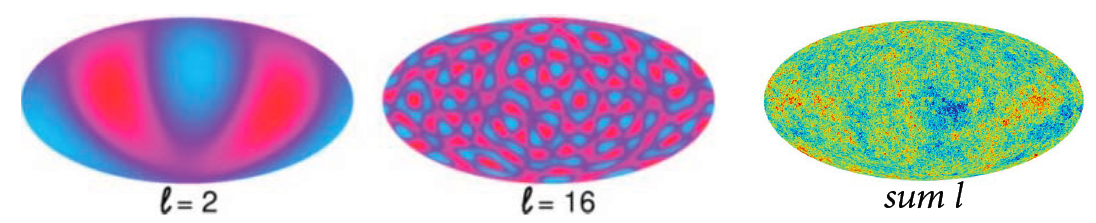} \\
\includegraphics[width=15cm]{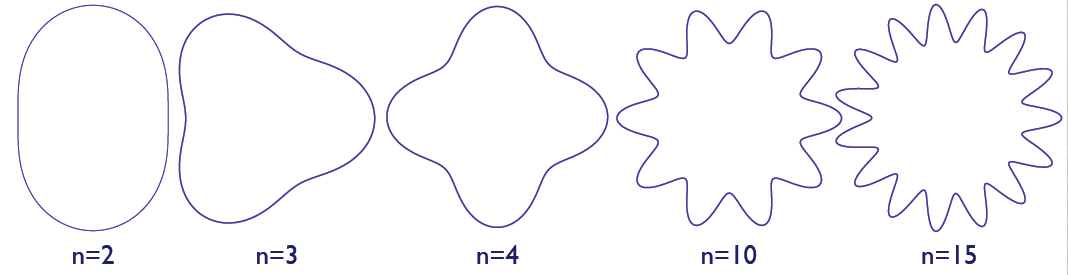}
\caption[]{ Top: the length scales in the CMB. Bottom: the length scales probed with higher $n$ in heavy-ion collisions.}
\label{fig:scales}
\vspace*{-0.1cm}
\end{centering}
\end{figure}

For the analogy with the CMB we are interested in temperature-correlations. Thus we need to consider how much of the initial inhomogeneity will be transferred into the final state.
Simulations based on hydrodynamic models indicate that collisions of Au nuclei (12 fm across) may contain hotspots of size 1.5 fm and that remnants of those hotspots persists during the collisions evolution~\cite{werner}. How much of this initial inhomogeneity is being transferred into the final state depends on the mean-free-path $l_{mfp}$  and it's relation to the size of the hotspots. The higher the value of $n$, the smaller the length scale we are probing. An $n=2$ asymmetry (think of $v_2$) can be translated because the $l_{mfp}$ is small compared to the length probed by $n=2$. That's why ideal hydro works well for $v_2$. The fact that hydrodynamic models do a reasonable job of predicting the value of $v_2$ suggests that $l_{mfp}$ can be considered small compared to the size of the system. But what about $n=3$ or even larger? As illustrated in Fig.~\ref{fig:scales}, as we increase $n$, we reduce the length scale probed. If the mean-free-path is greater than the length scale, there is no translation from coordinate-space into momentum-space. We only expect an efficient conversion of coordinate-space anisotropies into momentum-space when $l_{mfp}<2\pi\langle R\rangle/n$ where $\langle R\rangle$ is the average radial position of the systems constituents. This relation was checked in a simulation by calculating the initial participant eccentricity for all values of $n$ ($\varepsilon_{n,part}^2$)~\cite{v3} from a Monte Carlo Glauber model~\cite{mcg}. For central collisions, when the participants are treated as point like, $\varepsilon_{n,part}^2$ is nearly independent of $n$. When the participants are smeared over a region of size $r_{part}$, $\varepsilon_{n,part}^2$ at higher $n$ becomes quenched. We find that $\varepsilon_{n,part}^2$ is reduced by half when $r_{part}<2\pi\langle R\rangle/n$ and argue that $l_{mfp}$ will have a similar effect as $r_{part}$. The power-spectrum tells us the power transferred into each harmonic $n$ as a function of $n$. Clearly, the behavior of the power-spectrum can be used to extract information about length scales like $l_{mfp}$. By examining the power-spectrum of heavy-ion collisions which includes information for all values of $n$ (beyond just $n = 2$ or 3), we hope to better constrain $l_{mfp}$~\cite{slb}.  We note that other length scales that could be important have been identified: thermal scale; time of free streaming; size of hadrons; etc~\cite{shuryak}.

To determine the power-spectrum we take the STAR data of $p_T$ correlations vs the relative azimuthal angles and the relative rapidity between the particles~\cite{star}. We use this correlation measurement since this is a good proxy for point-to-point temperature-fluctuations: just like the blackbody spectrum, the $p_T$ spectra can reflect the temperature in heavy-ion collisions. A narrow peak positioned around small angle separation is observed in the data. This tells us that if a particle comes out with above average $p_T$, then the nearby particles also tend to have large $p_T$ . This is consistent with expectations from hotspots on the surface of last scattering. The correlation of these fast particles suggests that they are born out of the same high-density, high-temperature lump, i.e. the same hotspot. The relevance of temperature fluctuations to heavy-ion data is supported by the improvement to $p_T$ spectra fits using a Tsallis distribution~\cite{zhangbu}.

\begin{figure}
\includegraphics[width=16cm]{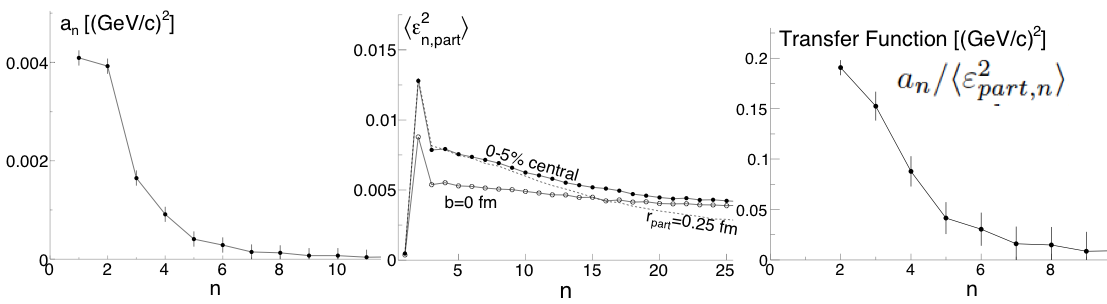}
\vspace*{-0.3cm}
\caption[]{Left: the power-spectrum $a_n$ extracted from $p_T$ correlations in heavy-ion collisions. Middle: the participant eccentricity calculated from a Monte-Carlo Glauber model. Right: the transfer-function for heavy-ion collisions showing the suppression at higher values of $n$ in the final momentum-space compared to the initial coordinate-space.
}
\label{fig:tfunc}
\vspace*{-0.1cm}
\end{figure}

We considered a slice of the $p_T$ correlations at zero relative rapidity vs relative azimuthal angle. We Fourier-transform this correlation to generate a power-spectrum vs harmonic number $n$. The coefficients of the Fourier decomposition $a_n$ give us the power-spectrum. This is shown in the left plot of Figure \ref{fig:tfunc}. The $a_n$ are similar to $v_n^2$ but extracted from $p_T$ correlations instead of number correlations~\cite{trainor}. $a_n$ is the power-spectrum for heavy-ion collisions and it is the analogous measurement to that of the CMB. We find that most of the power is in the lowest modes and $a_n$ falls off rapidly with $n$. Modes other than $n=2$ however still show a significant contribution.

For the CMB, the lowest modes are strongly suppressed and significant acoustic oscillations show up at $l=200$. This is due to the fact that in the early universe inflation stretches the fluctuations by 30 orders of magnitude so that the small wave numbers are outside the acoustic horizon. Mishra et. al had proposed looking for suppression of small $n$ modes due to an acoustic horizon in heavy-ion collisions~\cite{mishra}, but we see is clearly a suppression of large $n$ modes. We argue that this is because viscosity is far more important in heavy-ion collisions than in the early universe~\cite{knudsenfit}. 

Apparently, the damping due to various length-scales in heavy-ion collisions ($l_{mfp}$, etc.) is the dominant effect. Next we want to compare the final power-spectrum to the initial coordinate-space anisotropy. The azimuthal distribution of matter in the initial overlap zone is calculated using the participant eccentricity for all harmonics n $\varepsilon_{n,part}^2$~\cite{v3}. We start with a Glauber model~\cite{mcg} for the initial spatial distribution of participants for perfectly central collision (b=0 fm) in order to compare to data also for the 5 \% most central collisions. The results are shown in the middle plot of Figure \ref{fig:tfunc}. The large $n = 2$ term persists even for central collisions because we include the intrinsic deformation of the Au nucleus in our Monte Carlo. The $n = 1$ term is small because the participants are re-centered to the center of mass. For central collisions for n > 2 the eccentricity is nearly independent of $n$ (the higher harmonics are important too). This is because for symmetric collisions and point-like participants, $\varepsilon_{n,part}^2$ depends only on the number of participants, independent of n. We expect, however, that the conversion of higher harmonic eccentricity will be damped due to the existence of the length scale $l_{mfp}$. Giving the participants a size causes the eccentricity to fall off faster with $n$. The drop occurs where the size of the participant is larger than the lengthscale corresponding to $n$ the $\varepsilon_{n,part}^2$ is smeared out~\cite{sorensenhp}.

We define the ratio $a_n/\varepsilon_{n,part}^2$ as the transfer-function~\cite{slb} and show it on the right plot of Figure \ref{fig:tfunc}. The transfer-function gives us information about how efficiently the coordinate-space anisotropy has been converted into momentum-space. The transfer-function derived from data and the Glauber model shows that as $n$ increases, the efficiency for converting coordinate-space anisotropy into momentum-space quickly drops off. This can be expected from the condition that the transfer-function should go to zero when $l_{mfp}$ is larger than the length-scale associated with $n$:  $n>2\pi\langle R\rangle/l_{mfp}$. We can make a crude estimate for $l_{mfp}$ based on the transfer-function. If we take $\langle R\rangle = 3$ fm for the average radial position of a participant and n = 6 as the harmonic beyond which the conversion is inefficient, then we get $l_{mfp} = 3~$fm. This estimate can be improved with a more complete model of heavy-ion collisions. This estimate corresponds to a viscosity several times larger than estimates based on the centrality dependence of $v_2$~\cite{knudsenfit} or from the longitudinal width of $p_T$ correlations~\cite{gavinaziz}. Our crude estimate, however, is geometry based, not accounting for the various phases of the expansion.

We've used $p_T$ correlations from heavy-ion collisions to extract a power-spectrum analogous to the power-spectrum extracted from the Cosmic Microwave Background. We calculated the initial state eccentricity and then find the transfer-function which we define as the ratio of the power-spectrum to the initial eccentricity. We argue that the transfer-function required to describe the RHIC data may potentially be understood in terms of an inefficiency in conversion of coordinate-space anisotropy into momentum-space when $l_{mfp} > 2\pi\langle R\rangle/n$. We used this transfer-function to make a rough estimate of the mean-free-path of the system's constituents. The introduction of the harmonic axis $n$ provides more detailed information. This approach represents a new method for determining the characteristics of heavy-ion collisions and the QGP and should be further investigated. It stands as a challenge for models of heavy-ion collisions to correctly describe this transfer-function: in other words, to correctly describe the conversion of coordinate-space anisotropies at all scales into momentum-space anisotropies at all scales.


\end{document}